# Microrheology and structural quantification of hypercoagulable clots

L. Wolff-Trombini[1], A. Ceripa[2,3], J. Moreau[4], H. Galinat[5], C. James[1,6], N. Westbrook[4], J.M. Allain[2,3,*]

[1]Université de Bordeaux, UMR1034, Inserm, Biology of Cardiovascular Diseases, Pessac, France

[2]LMS, CNRS, Ecole Polytechnique, Institut Polytechnique de Paris, Palaiseau, France

[3]Inria, Palaiseau, France

[4]Université Paris-Saclay, Institut d'Optique Graduate School, CNRS, Laboratoire Charles Fabry, Palaiseau, France

[5]CHU de Brest, Service d'Hématologie Biologique, Brest, France

[6]CHU de Bordeaux, Laboratoire d'Hématologie, Pessac, France

*correspondance: jean-marc.allain@polytechnique.edu

## Abstract

We propose a combination of microrheological and structural characterizations of fibrin networks to study blood hypercoagulability. Fibrin is the central element of coagulation as its polymerization creates the network of fibers in which platelets and red blood cells are included. This is a controlled process via cascades between various coagulation factors. An alteration in the concentrations of coagulation factors and inhibitors will lead to hypocoagulation or hypercoagulation. These changes in the conditions of polymerization of fibrin lead to the formation of networks with different architectures and thus modify its mechanical behavior.

We have performed microrheology by recording Brownian motion of microbeads caught in the network of clots. The structure was quantified under the same polymerization conditions with confocal microscopy images. We have tested our approach by adding fibrinogen to the plasma, which leads to a stiffer, denser network with shorter fibers. The addition of coagulation Factor VIII at 400% induces the same correlated trend between a denser network and a higher modulus. This comparative approach is promising for the study of other conditions altering clot formation and may lead to a new diagnosis approach for hypercoagulability.

# INTRODUCTION

Clotting is the process by which the flow of blood is stopped after an injury. This process may occur pathologically, in both arteries and veins. In the venous system, the severe forms are called Venous Thromboembolic Events (VTE), which include Deep Vein Thrombosis (DVT) and Pulmonary Embolism (PE). Today, 50% of recurrent VTE remain unexplained.

Coagulation is induced by a cascade activation of coagulation factors. At the end of this cascade is the activation of thrombin, which catalyzes the polymerization of fibrinogen into fibrin. This forms a network of small diameter fibers, which gel the blood and block the flow. Polymerization conditions will modify the size of the fibers, their length and density, as well as the proportion of fibrinogen polymerized, and the speed of clot setting. This induces a modification of the organization of the fibrin network (1). This structure is usually observed directly either by scanning electron microscopy (SEM) or by confocal microscopy. SEM imaging has the advantage of being highly resolved, giving access to the diameter of the fibers, with a large depth of field providing a good overview of the structure. However, it requires complex sample preparation and provides only two-dimensional images. On the other hand, confocal microscopy has less risk of altering the network but requires the introduction of a fraction of fluorescently labeled fibrinogen. It also allows a reconstruction of the network in volume, albeit with a resolution not sufficient for a precise measurement of fiber diameters. Using these structural imaging methods, it has been shown that hemophilia is associated with a sparse network (2), while high thrombin concentrations favor a dense network (3). However, these methods alone sometimes give contradictory results: Zabczyk *et al.* show that patients with recurrent PEs have denser clots than those with non-recurrent PEs in SEM (4), while Baker *et al.* find no difference between recurrent and non-recurrent VTEs in confocal microscopy (5).

Another approach to observe coagulation-induced gelation is based on rheology (6, 7) as the network will strongly stiffen when it polymerizes. Rheological measurements are classically made with macroscopic rheometers. Clinically, thromboelastogram (TEG) and Rotational thromboelastometry (ROTEM) are commonly used to study the dynamics of clot formation and their resistance to fibrinolysis. These methods give a rapid diagnosis on the hemostatic situation of patients (8). In research, classical rheometers can be used, placing the clot between two metal plates before moving one of the two. Macroscopic rheology gives access to the low frequency behavior, at the scale of the whole network. Alternatively, microrheological methods have been developed, which probe the clot either actively or passively. Active methods, such as magnetic beads, have more or less the same advantages and limitations than macroscopic rheology but test the clot at a micrometric scale. Passive methods take advantage of the thermal fluctuations of beads or probes in the clot (9, 10) to determine the mechanical properties. They have the advantage of giving access to the high-frequency behavior (so high shear rate), with the drawback of being limited to the small shears. In this way, it has been shown that fibrin behaves as a semi-flexible polymer, with a transition between 1 Hz and 10 Hz (5, 10). It has also been possible to associate a decrease in storage modulus in clots from patients with recurrent VTE (5).

Microstructure and mechanical properties are linked in fiber networks, as explored in previous studies (1, 5, 10–12). However, no clear trend was observed, apart from a link between fibrin concentration and clot stiffness. This remains surprising, as an excess in a coagulation factor is likely to decrease the diameter of the fibers, reduce their length, increase the branchpoint density and lead to an increase in the network rigidity (1). A difficulty in the previous approaches was that the conditions of the clot formation were different for the microstructural and the rheological characterizations. We propose here a microrheology system using bead tracking by a reflected laser, in parallel with a structure measurement by confocal microscopy. The conditions of clot polymerization being similar, our measurements are directly comparable. We have used these two approaches to discriminate unmodified clots consisting of Platelet-Poor Plasma (PPP) from a pool of donors, from those to which known hypercoagulability factors (fibrinogen or Factor VIII) had been added.

## MATERIALS AND METHODS

### Protocol of clot preparation

Clots are made of Platelet-Poor Plasma (PPP), mixed with PPP Reagent (a mixture of tissue factor and phospholipids). As PPP is citrated to avoid coagulation, calcium has to be added. Beads are added for the microrheology measurements, and fluorescent fibrinogen for the confocal microscopy. All our clots include both, to make sure concentrations are identical for confocal and mechanical characterizations.

PPP was obtained from the blood of healthy adults. Blood was collected by venipuncture (vol/vol, 9:1 of 3.2% trisodium citrate), and centrifuged twice at $2500g$ for 15 min (agreement number: 20CADRE001). Supernatant was aliquoted and immediately frozen at -80°C. Individual PPP were pooled and used as PPP for all experiments. PPP was unfrozen at room temperature just prior to the experiment.

PPP reagent was obtained from PPP reagent "normal" (Stago, France), rehydrated with 500 µL of HEPES buffer, at least 30 min prior to the experiment.

The $CaCl_2$ solution was made by dissolving calcium chloride in a HEPES solution. The concentration of the solution is adjusted in each sample, to reach a final calcium concentration of 10 mM in the clot.

The choice of volumes for the preparation of the PPP reagent and for the $CaCl_2$ solution are driven by the need to have a proper coagulation, not too fast so that we can mix the clot. We targeted 5 pM of tissue factor as the final concentration, low enough so that the intrinsic amplification pathway of thrombin generation could be analyzed. Calcium concentration was chosen to be 10 mM, similar to concentrations found in the literature (13, 14).

We let the clot coagulate for 30 min at 37°C, to make sure thrombin generation, which peaks about 10 min after the addition of calcium, was completed. Even though the clot does not appear to evolve after this peak (3), we chose to add this extra coagulation time. However, as the chemical reactions are not stopped, the fibrinolysis will eventually dissolve the clot which limits the time for measurements. Using confocal microscopy, we observed the microstructure of the clot right after the 30 min of coagulation, then after 1 hour, 2 hours and 4 hours at room temperature. We did not observe any significant difference in the structure up to 2 hours after coagulation. After 4 hours, the fluorescence signal became diffuse, likely due to the beginning of the clot lyse. We thus did confocal and microrheology measurements within a limit of 2 hours.

**Standard protocol:**

The following protocol was used as a reference in our experiments:

200 µL of PPP was mixed with 10 µL of 5 mg/mL AlexaFluor 488-conjugated fibrinogen (Molecular Probes) and with 2 µL of 6 µm carboxylate polystyrene microspheres (2.1 $10^8$ particles/mL, Polybeads®).

A pit was made by gluing cloning cylinders (Bel-Art, FisherScientific) on a coverslip. The bottom of the pit was coated with 21 µL of PPP reagent, then 200 µL of the PPP mixture, preheated for 10 min at 37°C, was placed in the pit. 29 µL of a $CaCl_2$ solution at 86 mM was added to induce the coagulation. The mixture was gently stirred. The clot was left to coagulate for 30 min at 37°C. The pit was removed from the oven, and a microscope slide was glued on the cylinder to close it, so that it could not leak and was easy to manipulate. The clot was used in the next 2 hours.

**Fibrinogen-enriched clots:**

The protocol was the same one as for "standard" clots, except that 25 µL of human fibrinogen (Clottafact, LFB) was added to the PPP mixture. For the control, the fibrinogen was replaced by 25 µL of HEPES buffer. As before, 200 µL of the PPP mixture was placed in the pit. Finally, 4 µL of a $CaCl_2$ solution at 625 mM was added to induce the coagulation.

**Factor VIII-enriched clots:**

The protocol was the same one as for "standard" clots, except that 1.2 µL of recombinant porcine Factor VIII (Obizur, Takeda) was added to the PPP mixture. For the control, the Factor VIII was replaced by 1.2 µL of HEPES buffer. 200 µL of the PPP mixture was placed in the pit and 27.8 µL of a $CaCl_2$ solution at 90 mM was added to induce the coagulation. This led to a concentration of Factor VIII 4 times larger than the normal one (400% in Factor VIII).

Note that we chose to maintain the same concentrations of fibrinogen and coagulation factors for the control and the corresponding enriched clot. Consequently, there are slight differences between the different controls, and thus small variations of their measured parameters.

**Microrheology**

The mechanical properties at the microscopic scale were measured by passive microrheology using a detection system schematized in Figure 1. Compared to other passive rheology set-ups (10, 15, 16), it has the particularity of working in reflection. An Ytterbium fiber laser (Keopsys, λ = 1069 nm, TEM00, linearly polarized, 150 to 600 mW power) delivering 230 mW at the entrance of a microscope objective (Nikon PlanFluor, 100x, NA 1.3) was focused on a bead to measure its displacement. Through a polarizing beamsplitter (PBS), a quarter wave plate ($\lambda/4$) and a dichroic beamsplitter (DBS), the laser light reflected by the bead was detected on the quadrant photodiode (QPD, SPOT-9DMI, OSI Optoelectronics) to record the Brownian motion of the bead. Two cameras were used to guide the centering of the bead on the laser focus, using the XYZ piezo stage on which the sample is mounted, taking advantage of the reflection of the laser on the bead (see Fig. 2).

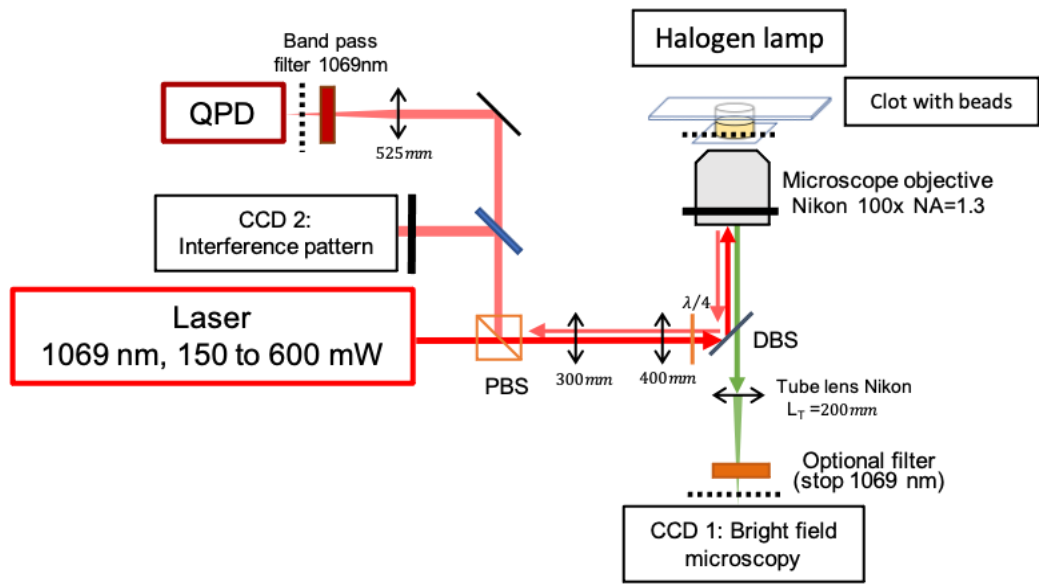

**Figure 1**: **Experimental setup of the detection system in reflection.** A laser at 1069nm is focused through a microscope objective on a bead inserted in the clot. The cameras CCD 1 and CCD 2 are used for centering the bead on the laser focus and the quadrant photodiode (QPD) for measuring the Brownian motion of the bead. Lenses with focal lengths 300 and 400nm form a telescope that expands the laser beam and conjugate CCD 2 with the back focal plane of the objective. Dotted and solid lines show two series of conjugate planes: the bead, CCD 1 and the QPD on one side, CCD 2 and the back focal plane of the objective on the other.

The first camera monitored the focusing on the bead. To do so, the clot is observed in bright field microscopy with the camera (see, in Figure 1, CCD 1 with filter). To choose the height of the measured beads, we used as a reference the reflection of the laser focus on the cover slip and then moved the sample down by a controlled amount with the piezo stage, looking for beads in focus in the range of 40 to 60 µm height from the coverslip. Difference in the appearance of the bead, as shown in Figure 2b, allows us to adjust the Z position of the bead with a precision of 1 µm. Finally, we used the laser light reflected on the bead (CCD 1 without filter) to center the bead on the laser by moving the piezo stage laterally in the X and Y directions.

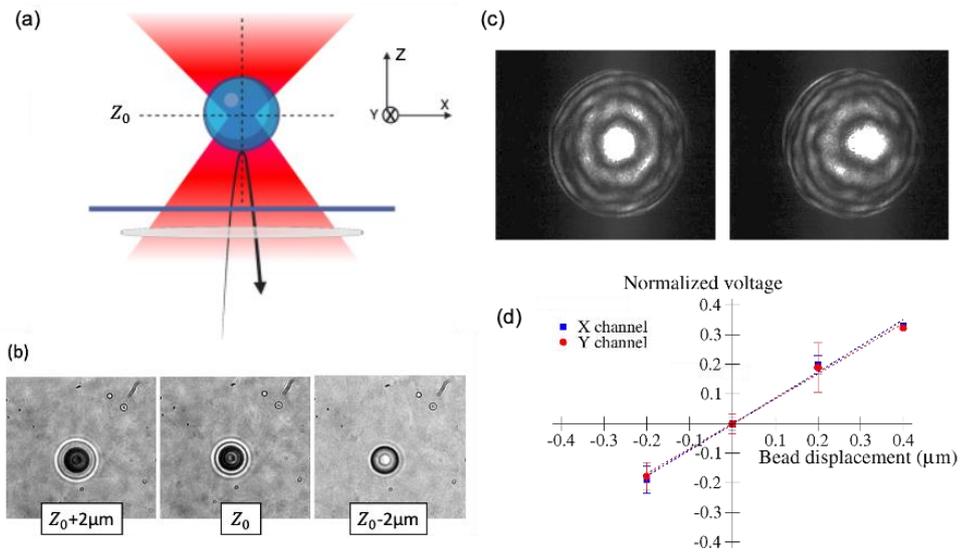

**Figure 2**: **Step-by-step method for centering the bead on the laser focus**.
(a) Schematics of the laser interacting with the bead in the centered position. (b) Image of the bead on CCD 1 when the bead is 2 µm below the focal plane ($Z_0$ + 2 µm), when the bead is in the focal plane ($Z_0$) and 2 µm above the focal plane ($Z_0$ - 2 µm). (c) Interference pattern on CCD 2 for a centered bead in X and Y (left) and for a bead displaced 200 nm in the X direction (right). (d) Normalized voltage measured by the QPD as a function of the displacement in X and Y (in µm) applied to the bead by the piezo. The slopes on this plot give the calibration factors of the QPD signal in X and Y in µm$^{-1}$.

A more precise lateral centering of the bead on the laser focus is achieved with the second camera (CCD 2). It records the interference pattern at infinity (in the back focal plane of the objective, conjugated with CCD 2, see Fig. 1) generated by the superposition of 3 reflected beams: on the coverslip and on the lower and upper surfaces of the bead. The symmetry of this interference pattern is very sensitive to the position of the bead with respect to the focus of the laser. We estimated the precision of this centering in X and Y to 200 nm (see Fig. 2c).

Finally, the normalized voltages on the QPD (difference of quadrants divided by their sum) must be converted into displacement in the plane of the bead. This calibration is performed by moving the piezo stage in 4 steps of 0.2 µm in X and Y, and recording the center of the Brownian motion. When the bead is well centered on the laser focus, the signal remains linear with the piezo displacement. The slope gives us a calibration factor for each Brownian motion recording: on average, 1.03 µm$^{-1}$ on X, and 0.98 µm$^{-1}$ on Y.

From the Brownian motion of a bead, we can calculate the local storage (G') and loss (G'') moduli of the medium surrounding the bead as a function of angular frequency (9).

Figure 3 describes the main steps of the procedure. First, the power spectral densities (PSD) of the bead motions in X and Y are determined, and filtered to reduce the noise. Second, the compliances of the medium are computed from the PSDs, enabling to determine the storage and loss moduli in each direction. Finally, the X and Y values are averaged to reduce the noise, as the Brownian motions are symmetrical. For all clots, the variations of the storage and loss moduli with frequency have similar behavior (see Fig. 3c). We chose to characterize the mechanical response of each bead at two frequencies, 30 and 3000 rad/s, one on each side of the transition frequency at 100 rad/s. To obtain these values, the storage and loss moduli are fitted by a power-law in the ranges 10-100 rad/s and $10^3$-$10^4$ rad/s. We recorded 3 to 4 Brownian motions per bead. The same analysis is performed on each Brownian motions and the associated storage and loss moduli are averaged; the mean values being considered as the moduli of this bead.

In all our experiments, we observed a correlation between the storage modulus at low frequency (30 rad/s) and the other moduli that we measured (loss moduli at 30 rad/s and 3000 rad/s, and storage modulus at 3000 rad/s). Consequently, we chose the storage modulus at 30 rad/s to be our reference measurement.

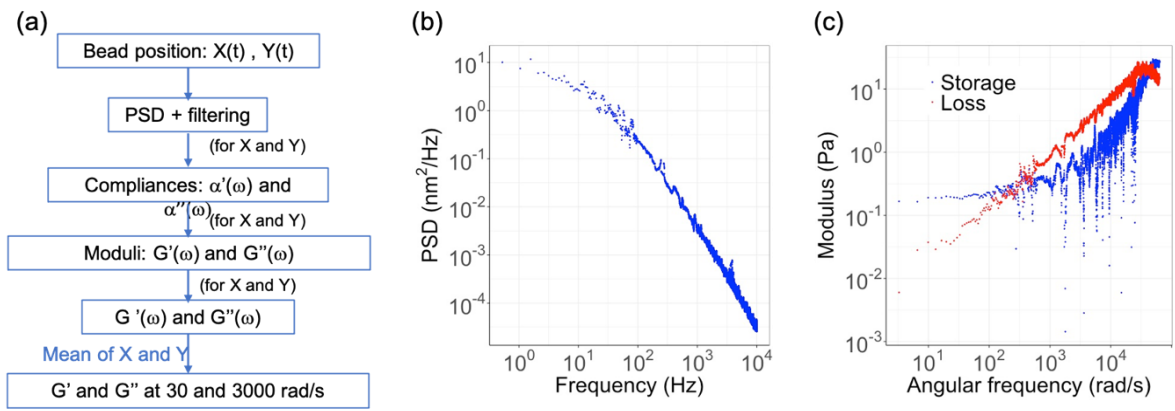

**Figure 3: Determination of the viscoelastic moduli.** (a) Principle of the moduli determination. Starting from the recorded bead positions, the PSDs in each direction were computed, and filtered to reduce the noise: subfigure (b) shows an example of a PSD graph in one direction after filtering. Then, the compliances were determined, followed by the moduli G' (storage) and G'' (loss) for each direction. Finally, we took the mean of the moduli in the two directions to obtain the moduli G' and G'': subfigure (c) shows an example of the moduli G' (storage) and G'' (loss) as a function of the angular frequency ω. Finally, to determine their values at 30 and 3000 rad/s, the moduli are fitted by a power-law around each of those two angular frequencies.

The moduli were measured with 6 μm beads. We also tested beads with 3 and 10 μm diameters but found that the dispersion in the measurements of storage modulus were larger. It could be due to the centering of the bead on the laser that was more critical with the 3 µm beads, and less precise for the 10 µm beads.

### 3-dimensional fibrin structure analysis

To determine the structure of the clots and its possible modifications, we imaged the clots with a confocal microscope (LSM 700, Zeiss). We imaged a stack of 160 µm by 160 µm by 21.5 µm (corresponding to 50 images with a voxel of 156 nm x 156 nm x 430 nm), starting at 55 µm from the coverslip, so that we are close to the height chosen for the microrheology measurement. To probe the homogeneity of the structure, we used a series of stacks starting at the coverslip and going up to 100 µm inside the sample, each stack having the same dimensions (160 µm by 160 µm by 21.5 µm).

For the quantitative analysis we used Fiji (doi:10.1038/nmeth.2019) to analyze each stack. Figure 4 presents the main steps of the procedure. Fibrin network parameters (fiber length and fibrin density) were quantified by performing several operations (threshold, filtering, skeletonization) on the fibrin stacks. Segmentation data were post-processed with a home-made algorithm (using Matlab software) in order to remove artifacts created by the skeletonization step, such as short artificial branches due to a thick structure (see Supporting Material for more details).

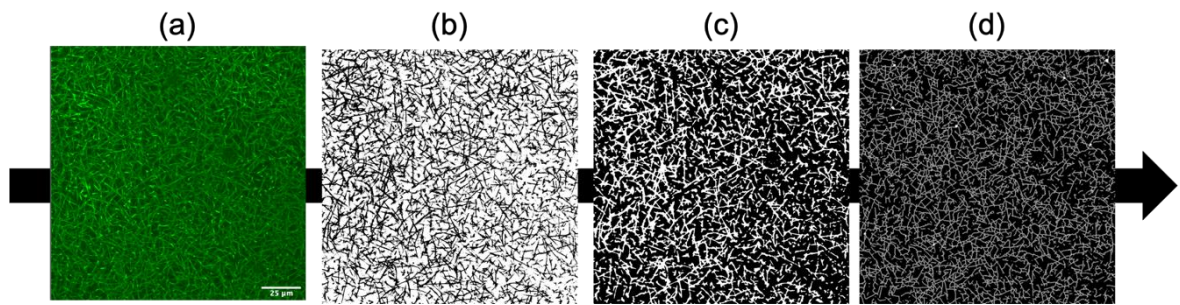

**Figure 4 – Three-dimensional (3D) quantification of a standard clot**. All images shown here are the Z-projection of 10 optical sections for clarity, but the quantification is done on the full Z-stack of 50 optical sections, recorded with a confocal microscope. Analysis was made using Fiji software up to the skeletonized image. (a) Raw micrograph of a fibrin network. (b) The fluorescence distribution is used to determine an optimal threshold, and thus a noisy binary image. (c) Filtered image. (d) The filtered image is skeletonized. A last step is performed to remove the artifacts (see Supp. Mat.) due to the skeletonization. Density of branching points, density of fibers, and length of the fibers are then determined.

### Statistics

Statistical tests were performed using Jamovi software. We used Mann–Whitney *U* test, with the assumption that one parameter was greater than another when a trend was clearly visible. Removing this assumption generally slightly affects the final p-value, without changing the conclusion. We used a non-parametric test as we are comparing small populations.

## RESULTS

### Effect of height on clot mechanics and structure

We first studied the effect of height on the modulus and structure of the clot. The modulus is given at 30 rad/s, measured by tracking the movement of 6 μm beads at different heights: 2 to 5 beads in each of the 4 different clots. The dependence with height was the same in all clots (see Fig. 5). We observed a decrease of the storage modulus with height, well described by an exponential decay over a typical height of 13 μm, before an asymptotic plateau at 0.1 Pa. This variation with height does not show in the structure of the clot itself. The density of fibers per unit volume does not vary with height, while the fiber length increases very slightly, by about 6% in 100 μm. This confirms the homogeneity of the clots prepared with our protocol. We also measured the structural parameters at different lateral locations in the sample, without observing any significant variation.

These results suggest that the stiffness of the network is increased by edge effects. This is why we selected in the following experiments beads at a height between 40 and 60 μm: high enough to reach the asymptotic value of the modulus, but not too high to limit scattering of the laser in the clot.

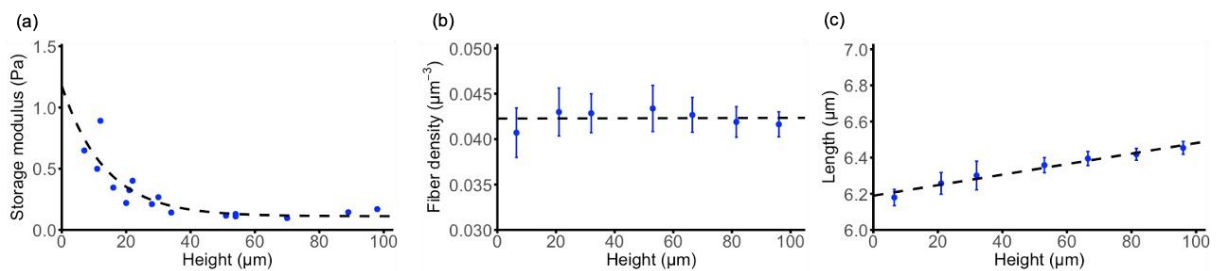

**Figure 5 – Height dependence of viscoelastic moduli and of structural parameters**
(a) Storage modulus measured for multiple beads at different heights in 4 normal clots (one dot represents one bead); (b) Fiber density (number of fibers per unit of volume) and (c) fiber length measured in normal clots at different heights. In (b) and (c), one dot represents the average measurement and its standard deviation for 4 clots.

### Effect of fibrinogen enrichment

To induce a modification of the network, we first enriched the plasma with 10% of fibrinogen. This should lead either to larger fibers or to a denser network. Both mechanisms can explain the increase in network stiffness that we observed (see Fig. 6a). As expected, the increase in fibrinogen concentration resulted in a significant increase in fiber density, by 50% and a decrease in the length of the fibers, by 1.6% (see Fig. 6b and c).

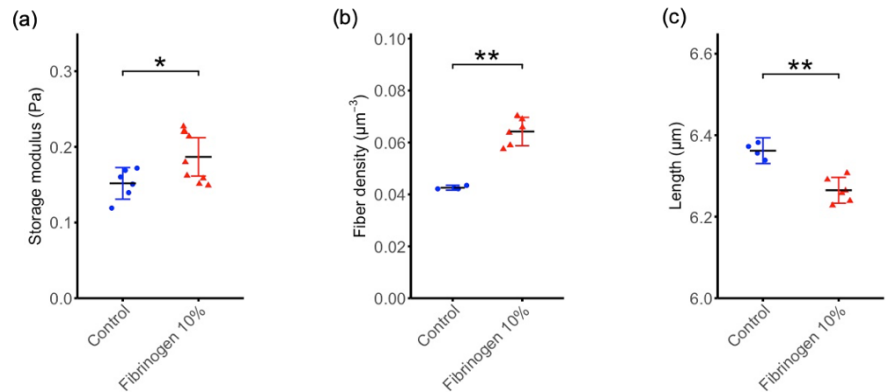

**Figure 6 – Mechano-structural properties of clots enriched in fibrinogen.**
(a) Storage modulus at 30 rad/s, (b) fiber density, and (c) fiber length, measured for control and clot enriched with 10% of fibrinogen. In (a), each point is the average value from three to five different beads in one given clot. In (b) and (c), each point is the measurement on a given clot. (*): p<0.05, (**): p<0.01.

### Effect of Factor VIII enrichment

To approximate a classical hypercoagulability situation, we added Factor VIII to the plasma, up to a concentration of 400% (4 times more than in clots obtained with control plasmas). This induced an increase in clot stiffness (see Fig. 7a), by 25%. We also observed an increase in fiber density and a decrease in fiber length (see Fig. 7b and c), by 19% and 1.3%, respectively.

We thus observed that an increase in polymerization factors favors the appearance of a larger number of shorter fibers, which is explained by a faster induction of polymerization.

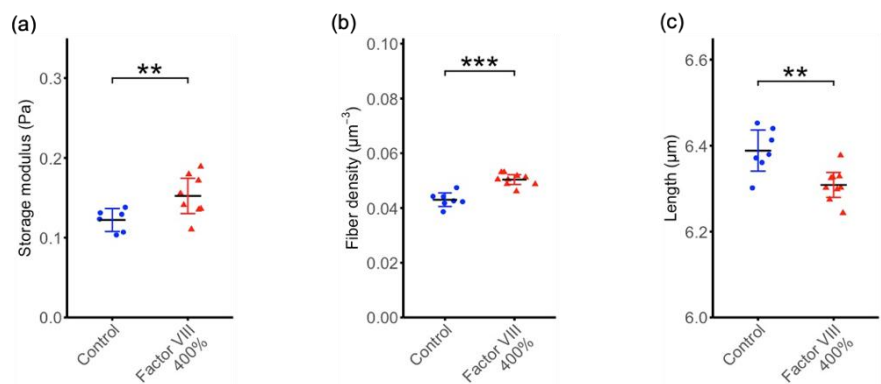

**Figure 7 – Mechano-structural properties of clots enriched in Factor VIII at 400%.**
(a) Storage modulus at 30 rad/s, (b) fiber density, and (c) fiber length, measured for control and clot enriched with 10% of coagulation Factor VIII. In (a), each point is the average value from three to five different beads in one given clot. In (b) and (c), each point is the measurement on a given clot. (**): p<0.01, (***): p<0.001.

## DISCUSSION

We set up a dual approach, allowing measurements of the rheology and of the microstructure of a clot in the same experimental conditions. We could not perform the two measurements on the same clot, as it evolves with time. However, we could quantify both properties for clots formed under the same conditions from PPP and observe the effect of factors mimicking hypercoagulability.

We developed a quantification of the fibrin network, with the determination of the length between branches, and the fibrin density. These parameters are obtained directly from the confocal images, after thresholding and removal of spurious branches. Another important parameter is the diameter of the fibers, but it cannot be measured by confocal microscopy, as it is around 100 nm (5, 17), smaller than the resolution limit.

The microrheological behavior of the fibrin network is similar to those reported in the literature (10, 16) (see Fig. 3c). At low frequencies, the storage modulus is constant, around a few tenths of Pa for normal clots. This modulus is in the wide range of values reported in the literature. It is superior to the loss modulus, which varies as a power-law of the angular frequency $\omega$, reflecting a solid-like behavior. At high frequencies, both loss and storage moduli vary as a power-law of $\omega$, with the loss modulus higher than the storage one, denoting a fluid-like response. The transition frequency between the two regimes was found in our experiments around 100 rad/s. At high frequency, the moduli vary as $\omega$ to the power 0.84. This is closer to 7/8 than to 3/4, which is the expected exponent for semi-flexible polymer networks (5, 10, 18). The exponent 7/8 may be associated with the fluctuations of the filaments in their longitudinal directions, although this was expected to play a minor role in the network response (19). We consider that this comes likely from the fact that we use a full plasma clot and not a reconstructed fibrin network: the clot microstructural organization depends strongly of the polymerization condition, which are different for plasma with all the coagulation factors than for solutions of only thrombin and fibrin. A similar exponent has been reported on fibrin networks made from human plasma (16).

The symmetry of the Brownian motion shows that the bead is in a quasi-isotropic medium, and thus that we do not measure the response of an individual fiber. Note that we did not observe any trapping of the bead by the detection laser: the displacement of the laser did not lead to any movement of the bead, even when the laser power was increased. We thus neglected the trap stiffness in the mechanical response of the bead. Our protocol enables us to prepare spatially-homogeneous clots (see Fig. 5). However, if we are too close from the glass surface (less than 40 µm), we observed an increase of the moduli (see Fig. 5a). This boundary effect is well known for the loss modulus (20).

Our goal was to evaluate whether hypercoagulability due to imbalances in coagulation factors was reflected in clot properties. To do this, we started by the addition of fibrinogen. Not surprisingly, this resulted in an increase in fibrin density (see Fig. 6b). As more fibrin promotes a tighter network, we expect shorter fiber lengths. However, the measured reduction in length is small compared to the increase in density. The excess of fibrin may translate into thicker fibers. However, it is also possible that tiny fibrils are created, with a size smaller than our resolution. In both cases, the logical consequence is that the network is stiffer, resulting in an increase in storage modulus (see Fig. 6a). This is a statistical result: if the average modulus increases, the dispersion also increases, with moduli sometimes equal to those of the control clots. This increase in dispersion is difficult to explain but may reflect a greater heterogeneity of the network, with pores that are larger or smaller than the size of the bead.

To mimic a known situation of hypercoagulability, we chose to add Factor VIII up to concentrations 4 times larger than the normal one (400%) (21–23). In this situation we observed an increase in the density of the network, which is explained by a greater activation of thrombin, which will transform more fibrinogen into fibrin (see Fig. 7b). This densification is less important than when fibrinogen is added directly. We observed a moderate but significant decrease in the length of the fibers, similar to that observed when fibrinogen was added (see Fig. 7c). This suggests that the diameter of the fiber increases, but less than with the addition of fibrin. This leads to an unsurprising increase in storage modulus (Fig. 7a). It should be noted that the absolute values of the moduli in the Factor VIII enrichment experiment are not directly comparable to those measured in the fibrinogen enrichment experiment. Indeed, the changes in control preparation protocols lead to slightly lower moduli for the control. However, the increase in storage modulus is similar with the addition of Factor VIII than with the enrichment in fibrinogen, but more discriminant (smaller p-value).

## CONCLUSION

Our approach enables us to observe both the structure and the microrheology of fibrin clots under the exact same polymerization condition, on plasma pooled from donors. We changed the polymerization conditions first by adding an excess of fibrinogen, and second by increasing the concentration of Factor VIII (by 400%). In both cases, we observed an increase of the fiber density and of the storage modulus, while the fiber length appears mainly unaffected, suggesting that the fiber diameter has probably increased. To go further, it would be interesting to explore different concentrations of coagulation factors, inhibitors, and fibrinogen, keeping in mind that the coagulation reactions, in cascade, do not lead to linear effects in the concentration of the different factors. The results of these mechano-structural measurements are encouraging towards a diagnostic test of thrombosis predisposition in patients.

## AUTHOR CONTRIBUTIONS

HG, CJ, NW and JMA designed research, LWT, AC, NW performed research, AC, JM, JMA contributed analytic tools, LWT, AC, JMA analyzed data, LW, NW, JMA wrote the manuscript.

## DECLARATION OF INTEREST

The authors declare no competing interests.

## ACKNOWLEDGEMENTS

This work was partly supported by Agence Nationale de la Recherche (contract ANR-11-EQPX-0029 Morphoscope2). The authors would like to thanks Dr. Christine Mouton and Ms. Françoise Rat for providing the Clottafact, Léo Viallet and Oscar Boyadjian-Braize for their help in the design of the microrheology experiment, and Pierre Mahou for his help on the confocal microscopy.

# Supporting Material

## Microrheology and structural quantification of hypercoagulable clots

L. Wolff-Trombini, A. Ceripa, J. Moreau, H. Galinat, C. James, N. Westbrook, J.M. Allain

### Cleaning of the skeletonized confocal images

After recovering the structural data (from the skeletonization) with Fiji, we observe extremely short fibers (see Fig. S1). These fibers do not correspond to real fibers, but are due to erroneous branchpoints. They can be created by a thick intersection of multiple fibers, which leads to multiple short branches during skeletonization (in the blue circle of Figure S1). Thick fibers can also lead to very short fibers, almost perpendicular to the main one (in the red circle of Figure S1).

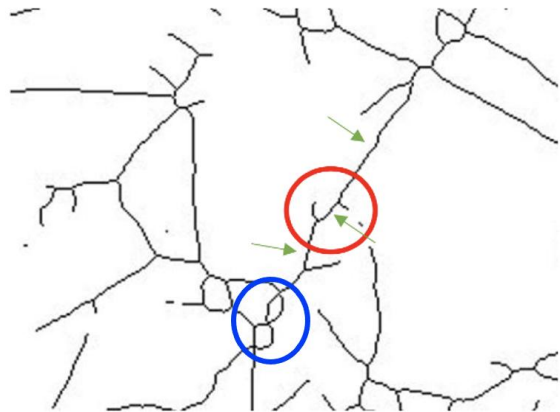

**Figure S1 - Artificial branches created after skeletonisation.**
Micro-fibers and branchpoints creation during skeletonisation. Green arrows show three split fibers, which should be 1 fiber, because of the surrounding microfibers (surrounded in red). Blue circle shows the non-existent branchpoints creating during the intersection of several fibers.

The first case detects more branchpoints than necessary at an intersection, and creates several small non-existent branches. The second case cuts long fibers into several sections, strongly reducing the length of the fiber. It also creates short fibers. Thus, they increase artificially the branching density, the number of fibers, and decrease artificially the fibers length.

To clean the image, our first step is to remove the terminal micro-branches. To do so, we define a cutoff length, which is the typical width of the fluorescence signal of a fiber. We consider that any segment shorter than this length is a spurious branch, and it will be deleted. If the segment is a terminal segment, we delete the segment and the only associated branchpoint. The two remaining segments that were also connected to this branch point are then merged into one larger segment (see Fig. S2, green circle).

The second step is to suppress the artificial branches at the intersection. To do so, we calculate the Euclidean distance between branchpoints. We consider that branchpoints which are closer than twice the cutting length are in fact a single intersection. Thus, we merge them, by creating a single branchpoint at their barycenter. Segments that were connected to the two deleted branchpoints are in turn deleted (see Fig. S2, red circle).

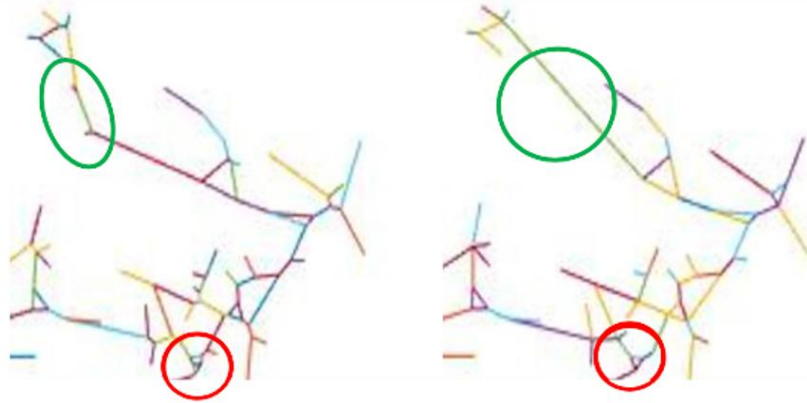

**Figure S2 - Skeletonized image before and after corrections.**
Left: an area of the skeletonized image, prior correction. Right: The same area of the skeleton. The green circle shows a long fiber artificial cut by erroneous micro-branches. The red circle shows the correction of an intersection region.

On a global scale, the skeleton keeps a similar appearance, while on a small scale, as we can see in Figure S2, some artifacts were corrected. This strongly modifies the length distribution.

This procedure leaves some artifacts. However, trying to correct too much the image pushes the other way around, with the suppression of short, visible, branches.